\documentclass[twocolumn]{aastex63}

\usepackage{gensymb}
\usepackage{amsmath}
\usepackage{threeparttable}

\shorttitle{PSR J1709's Proper Motion}
\shortauthors{de Vries et al.}

\begin{document}

\title{PSR J1709-4429's Proper Motion and its Relationship to SNR G343.1$-$2.3}

\correspondingauthor{Martijn de Vries}
\email{mndvries@stanford.edu}

\author{Martijn de Vries}
\affiliation{Department of Physics/KIPAC, 
Stanford University, Stanford, CA 94305-4060, USA}
\author{Roger W. Romani}
\affiliation{Department of Physics/KIPAC, 
Stanford University, Stanford, CA 94305-4060, USA}
\author{Oleg Kargaltsev}
\affiliation{Department of Physics, George Washington University, Washington, DC 20052, USA}
\author{George Pavlov}
\affiliation{Department of Astronomy \& Astrophysics, Pennsylvania State University, University Park, PA 16802, USA}
\author{Bettina Posselt}
\affiliation{Department of Astronomy \& Astrophysics, Pennsylvania State University, University Park, PA 16802, USA}
\affiliation{Department of Physics, University of Oxford, OX1 3PU Oxford, UK}
\author{Patrick Slane}
\affiliation{Harvard-Smithsonian Center for Astrophysics, 60 Garden Street, Cambridge, MA 02138, USA}
\author{Niccolo' Bucciantini}
\affiliation{INAF - Osservatorio Astrofisico di Arcetri, L.go Fermi 5, 50125, Firenze, Italy}
\author{C.-Y. Ng}
\affiliation{Department of Physics, The University of Hong Kong, Pokfulam Road, Hong Kong}
\author{Noel Klingler}
\affiliation{Department of Astronomy and Astrophysics, The Pennsylvania State University, 525 Davey Laboratory, University Park, PA 16802, USA}

\begin{abstract}
We have obtained a deep (670\,ks) {\it CXO} ACIS image of the remarkable pulsar wind nebula (PWN) of PSR J1709$-$4429, in 4 epochs during 2018-2019. Comparison with an archival 2004 data set provides a pulsar proper motion $\mu = 13 \pm 3$\,mas\,yr$^{-1}$ at a PA of $86 \pm 9 \degree$ ($1\sigma$ combined statistical and systematic uncertainties), precluding birth near the center of SNR G343.1$-$2.3. At the pulsar's characteristic age of 17\,kyr, the association can be preserved through a combination of progenitor wind, birth kick and PWN outflow. Associated TeV emission may, however, indicate explosion in an earlier supernova. Inter-epoch comparison of the X-ray images shows that the PWN is dynamic, but we are unable to conclusively measure flow speeds from blob motion. The pulsar has generated a radio/X-ray wind bubble, and we argue that the PWN's long narrow jets are swept back by shocked pulsar wind venting from this cavity. These jets may trace the polar magnetic field lines of the PWN flow, an interesting challenge for numerical modeling. 
\bigskip
\end{abstract}

\keywords{stars: neutron — pulsars: individual (PSR J1709$-$4429)}

\bigskip

\section{Introduction} 
\label{sec:intro}
PSR J1709-4429 is a young ($\tau_c = P/2\dot{P} = 1.7 \times 10^{4}$\,yr) pulsar, with spindown luminosity $\dot{E} \approx 4 \times 10^{36}$\,erg\,s$^{-1}$, originally discovered by \citet{Johnston1992}. At high energies, it was first observed with EGRET, making it one of the few known $\gamma$-ray pulsars at the time \citep{McAdam1993}. \textit{CXO} observations in 2000 and 2004 revealed X-ray pulsations \citep{Gotthelf2002}. The excellent angular resolution of \textit{CXO} also revealed the complex X-ray structure of the torus, equatorial outflow, and jets that appear to be swept back by the PSR motion \citep{Romani2005}. Fits to the torus and jets following the fitting procedure laid out by \citet{Ng2004}, yielded strong constraints on the position angle ($163.6 \pm 0.7 \degree$), inclination ($53_{-1.6}^{+1.4 \degree} $), and torus Doppler boosting factor ($\beta=0.70 \pm 0.01$).

The nearby supernova remnant (SNR) G343.1$-$2.3, which is observed at radio wavelengths, has long been discussed as the possible PSR birthsite. The objects appear to be at similar distances: the SNR distance was estimated from the rather rough $\Sigma-D$ relationship to be $\sim 3$\,kpc \citep{McAdam1993}, while the dispersion measure (DM) distance of the PSR, according to the NE2001 model of \citet{Cordes2002}, is $d \approx 2.3$\,kpc. However, radio scintillation measurements by \citet {Johnston1998} suggest a low transverse velocity $\approx 90$\,km\,s$^{-1}$. At the spindown age of J1709, this velocity is too low by a factor of a few for the PSR to originate from the current center of the SNR. 

More recently, observations with the the High Energy Stereoscopic System (H.E.S.S.) have shown an extended source of TeV emission west of the PSR  \citep{HESS2011}. This TeV emission could be interpreted as inverse-Compton emission from electrons deposited during the pulsars initial period of rapid spindown, cosmic rays generated by the SNR interacting with the surrounding interstellar medium, or some combination of both. 

Measuring the direction and size of the proper motion vector should provide an important clue of whether the PSR, SNR, and TeV source can be associated. However, this measurement has proven challenging with past observations. \citet{Romani2005} attempted to measure the proper motion using the \textit{CXO} observations from 2000 and 2004, but these data proved to be insufficient for a proper motion detection. Radio VLBI astrometry has also been attempted, but proved infeasible with the Australian LBA, as there is no suitable in-beam reference source. 

\begin{figure}[b]
\centering
\includegraphics[width=3.5 in]{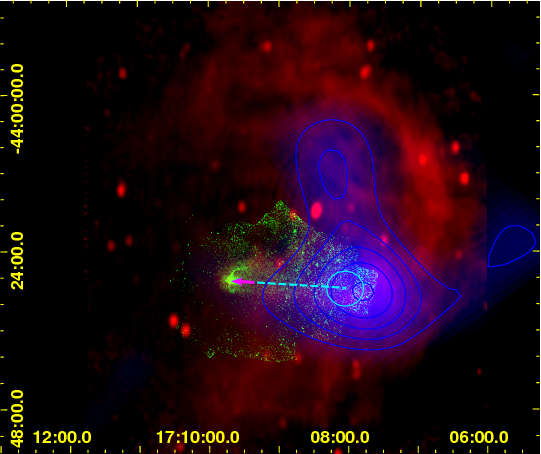} \\
\caption{Composite color image of J1709-4429 and the nearby SNR. Green: Chandra ACIS exposure-corrected 0.5-7 keV mosaic of all observations, smoothed (low exposure at the edges of the combined ACIS fields lead to poorly smoothed calibration artefacts). Blue w/ contours: H.E.S.S. TeV Galactic Plane Survey emission. Red: ATCA L-band (1.4 GHz) mosaic of SNR G343.1$-$2.3. The magenta arrow represents the PSR motion over its $1.7 \times 10^4$\,yr spindown age (see Section \ref{sec:pm}); the cyan extension shows $5\times$ this distance with an ellipse showing the $1\sigma$ proper motion uncertainty over this larger time span.}
\label{fig:largescale}
\end{figure}

We have obtained $670$\,ks of new \textit{CXO} observations between June 2018 and 2019. Combined with the archival observation from 2004, this gives us a $14.4$ year baseline allowing for direct astrometric measurement of the proper motion vector. The new data also show in more detail the pulsar wind bubble in which J1709 is embedded, and provide new constraints on the spectral properties of the emission in the PSR, torus, and jets.

\section{Observations and Data reduction}
\label{sec:data}

\begin{table}

\begin{flushleft}

\caption{Overview of J1709 \textit{CXO} observations used in this paper}
\begin{tabular}{ c  c c c c}
\hline \hline
Start Date  &  ObsID & Instrument & Exp & $\Theta$ \\
\,[yyyy-mm-dd] &  & & [ks] & [arcmin] \\ \hline
2000-08-14 & 757 & ACIS-S & 14.2 & $0^{\prime}.4$\\
2004-02-01 & 4608 & ACIS-I & 98.8 & $0^{\prime}.6$ \\ 
2018-06-25 & 20300 & ACIS-S & 55.3 & $1^{\prime}.0$\\ 
2018-06-26 & 20299 & ACIS-S & 113.3 &$0^{\prime}.8$\\
2018-06-29 & 21109 & ACIS-S & 60.2 & $1^{\prime}.0$\\
2018-10-08 & 20302 & ACIS-I & 51.1 & $2^{\prime}.4$\\
2018-10-10 & 21871 & ACIS-I & 2.8  & $2^{\prime}.4$ \\
2019-01-24 & 20303 & ACIS-I & 37.1 & $2^{\prime}.6$\\
2019-01-25 & 22060 & ACIS-I & 35.1 & $2^{\prime}.6$\\
2019-01-28 & 21899 & ACIS-I & 32.0 & $2^{\prime}.6$\\
2019-02-03 & 22075 & ACIS-I & 24.3 & $2^{\prime}.6$\\
2019-02-09 & 21869 & ACIS-I & 44.2 & $2^{\prime}.6$\\
2019-05-23 & 20301 & ACIS-I & 34.0 & $3^{\prime}.0$\\
2019-05-24 & 21870 & ACIS-I & 19.8 & $3^{\prime}.0$\\
2019-06-03 & 22244 & ACIS-I & 29.7 & $2^{\prime}.9$\\
2019-06-04 & 22005 & ACIS-I & 29.6 & $3^{\prime}.0$\\
2019-06-06 & 22246 & ACIS-I & 27.8 & $2^{\prime}.9$\\
2019-06-06 & 20304 & ACIS-I & 27.8 & $3^{\prime}.0$\\
2019-06-09 & 22245 & ACIS-I & 27.9 & $2^{\prime}.9$\\
2019-06-16 & 22087 & ACIS-I & 19.8 & $3^{\prime}.0$\\
\hline
 Total & & & 784.8 \\ \hline
\end{tabular} \\
Note - $\Theta$ indicates the distance in arcmin between the PSR and the on-axis position of the telescope.
\label{tab:ChandraObs}

\end{flushleft}

\end{table}

\begin{figure}

\centering
\includegraphics[width=3.2in]{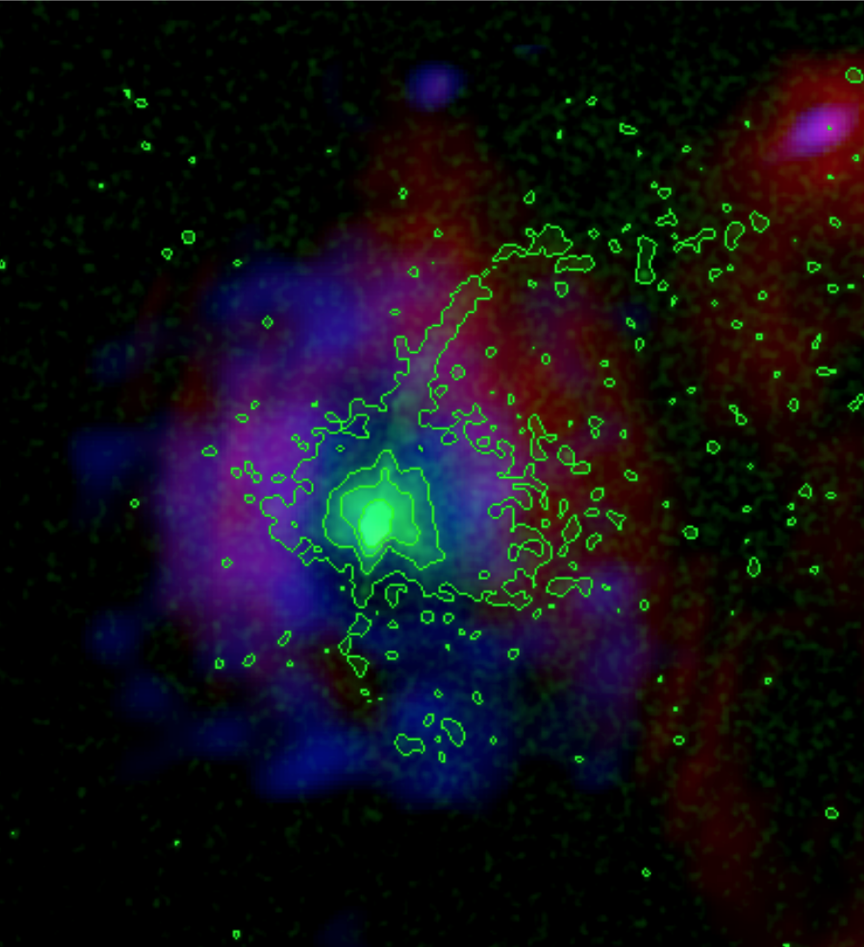} \\
\caption{$6^{\prime}$x$6^{\prime}$ composite color image of the pulsar, outflow, and jets. Red: ATCA L-band (1.4 GHz). Green: Chandra ACIS exposure-corrected 0.5-7 keV image, smoothed by a  $2.5^{\prime\prime}$ Gaussian. Blue: ATCA C-band (4.8 GHz)}
\label{fig:psroutflow}
\end{figure}

\begin{figure}
\centering
\includegraphics[width=3.4in]{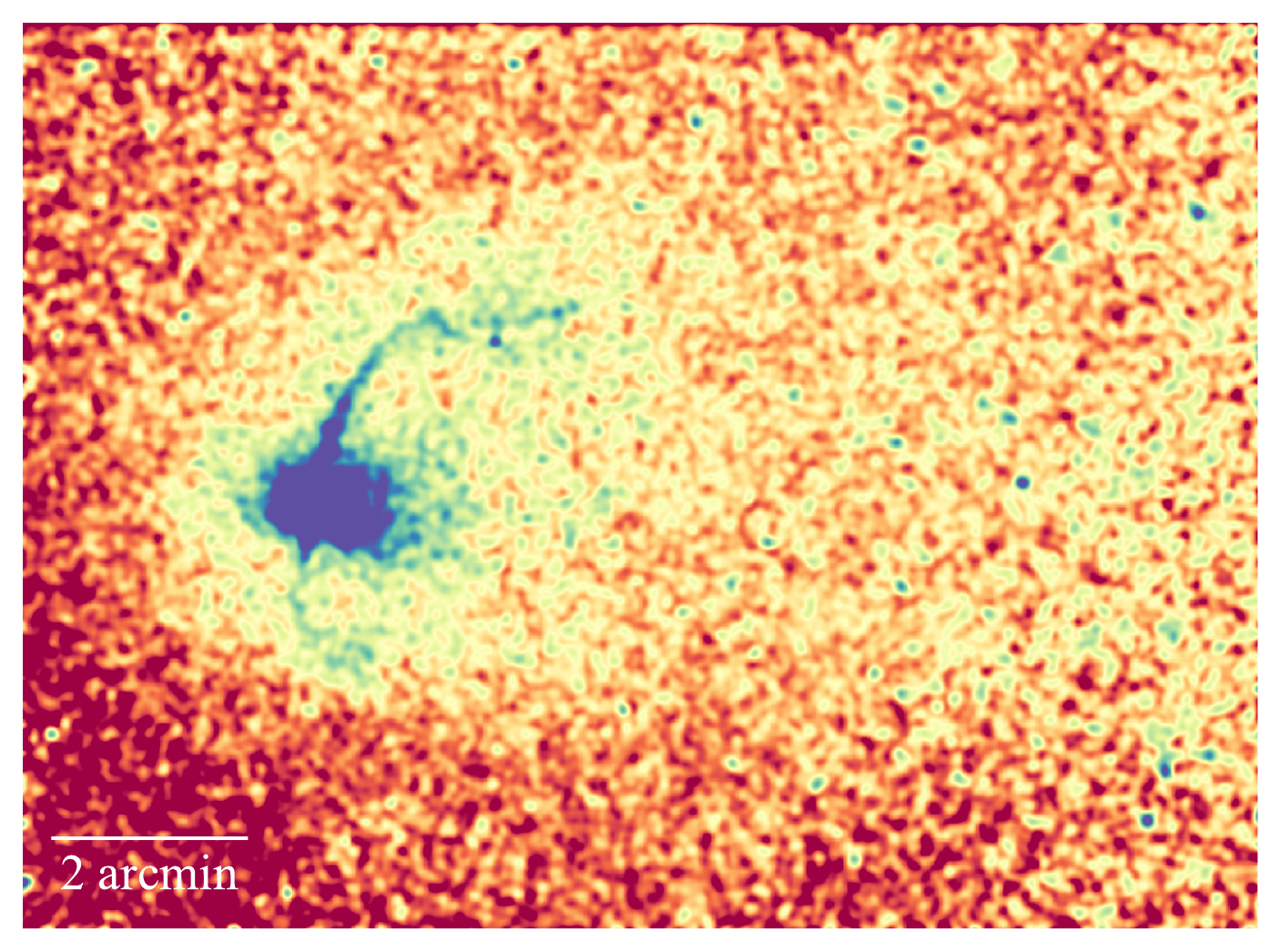} \\
\caption{Exposure-corrected, $0.5$-$4$\,keV image of J1709, smoothed by a $2.5^{\prime\prime}$ Gaussian. The stretch is chosen to highlight the edge of the wind bubble on the eastern side and the diffuse emission trailing the PSR on the western side. }
\label{fig:bowshock}
\end{figure}

J1709 was observed with \textit{CXO} for a total of 670\,ks between June 2018 and June 2019. Three of the observations in June 2018 were made with the ACIS-S array, while the rest use ACIS-I. Additionally, there are two archival observations, 15\,ks in 2000 and 99\,ks in 2004, described in more detail in \citet{Romani2005}. A full overview of the observations is given in Table \ref{tab:ChandraObs}.

The data were processed with the standard CIAO reprocessing tools, using CIAO 4.12 and CALDB 4.9.1. An astrometric correction was applied to the data, by cutting out a $2^{\prime}\times2^{\prime}$ region centered on the PSR and outflow in each of the observations. We used ObsID 04608 as a reference observation because of its long exposure time, and reprojected the other ObsIDs onto the sky frame of the reference observation. Each of the ObsIDs were cross-correlated with the reference observation by shifting them in RA and DEC directions. The cross-correlation was then fit with a Lorentzian function to determine the coordinate offset, which was applied to the observation with the CIAO tool \textit{wcs\_update}. With this astrometry correction, all observations are approximately aligned on the PSR.

We created ACIS `blank-sky' background maps for each CCD of each observation, scaling them by the number of 10-12\,keV counts of that observation. We also created instrument and exposure maps using the CIAO tools \textit{mkinstmap} and \textit{mkexpmap}, using an input spectrum  extracted from the PWN and PSR as a whole. Using the event data, blank-sky backgrounds, and exposure maps, we created a background-subtracted, point source-removed, exposure-corrected mosaic image by stacking all the observations.

Figure \ref{fig:largescale} shows the exposure-corrected X-ray image (using only pixels with 10\% or more of the maximum exposure, green), overlaid with ATCA L-band  \citep[red,][]{Dodson2002}, and H.E.S.S. TeV Galactic plane survey data \citep[blue,][]{HESS2018}. An arc of supernova remnant G343.1$-$2.3 is visible in the L-band data, while a ridge of radio emission extends from west to east, reaching the pulsar. The X-ray pulsar wind nebula shows the equatorial torus and swept-back jets, along with a faint trail of diffuse X-ray emission along the radio ridge. The H.E.S.S. emission shows a peak west of J1709. This is extended toward the pulsar and a peak of emission lies to the north, within the arc of the radio shell. The faint TeV emission to the west of the radio shell may be Galactic plane background. Note that the radio emission, the TeV extension and the low surface brightness X-ray trail all coincide as a horizontal band projected across the face of the SNR arc, pointing eastwards towards J1709. We will refer to this as the pulsar wind trail.

Figure \ref{fig:psroutflow} shows in more detail the interaction between the outflows and the radio PWN. The radio PWN is hollow, surrounding the X-ray equatorial flow, with channels corresponding to the jets, which are swept back on arcminute scales. 
Figure \ref{fig:bowshock} is stretched to emphasize the jet sweep back and the faint diffuse \textit{CXO} emission in an east-west band. This emission ends $\theta_0 \approx 1.5^{\prime}$ to the east of the PSR. The overall morphology suggest an eastward pulsar velocity across this horizontally extended wind bubble with the leading edge well in front of the pulsar and the X-ray jets swept back inside. A detailed measurement of the pulsar motion can check this scenario. 

\section{Source Registration and Proper motion}
\label{sec:pm}

Given the low scintillation velocity, the proper motion shift of J1709 is expected to be a fraction of an ACIS pixel ($0.492^{\prime\prime}$), even over the comparatively long baseline spanned by our data. To measure this proper motion as accurately as possible, we register the frames in each frame using the `Figure of Merit' technique \citep[`FoM'][]{Etten2012, Auchettl2015, deVries2020}. This technique compares the counts of observed registration sources on a sub-pixel grid, to predicted counts from simulated PSF models of those sources as a function of point source offset, determining the source position (relative to a selected reference frame) to sub-pixel accuracy. We elaborate on this technique below.

One goal with the 2018-2019 ObsIDs was to cover the faint diffuse emission (and to keep the jets within read-out node boundaries). Thus in most ObsIDs, the pulsar is $>2^{\prime}.4$ away from the optical axis (Table \ref{tab:ChandraObs}). This led to significant PSF deformation even for the pulsar and nearby point sources. We therefore restrict our proper motion analysis to four frames in which the PSR is the closest to the optical axis: the reference frame, ObsID 4608, and 3 frames taken in late June 2018: ObsIDs 20299, 20300, and 21109. This gives us a 14.4 year baseline over which to measure the proper motion of the pulsar.

\subsection{Frame offsets}

We selected a set of registration sources based on three criteria: 1) the sources are visible in the reference frame and in at least two of the 2018 frames, 2) they do not display significant variability between the observations, and 3) they have at least $\sim 10$ counts between $0.5-3.5$\,keV per observation. Unfortunately, ObsIDs 20299 and 20300 only use 3 chips (S2, S3, and S4) and ObsID 21109 uses 2 chips (S3 and S4), limiting the number of available registration sources. We have selected a set of 14 registration sources, all of which are in the field of view in ObsIDs 04608, 20299, and 20300. For ObsID 21109, only 11 of the 14 sources fall on the ACIS chips. The sources are on average $\sim 2^{\prime}.5$ away from the PSR, with distances ranging from $50"$ to $6'$. For each source, we looked up the hardness ratio in the Chandra Source Catalog 2.0, as well as potential optical counterparts in the Gaia DR2 catalog. Five of the sources have soft X-ray spectra, as well as an optical Gaia counterpart with known proper motion and are thus likely coronal emission from nearby field stars. We correct for the proper motions of these objects in the analysis below. The remaining nine sources have hard X-ray spectra and no detected optical counterpart. We conclude that these sources are likely to be extragalactic in origin and therefore have negligible proper motions.

For each source, we determined the reference positions in the reference frame using the CIAO tool \textit{wavdetect}, and created spectra by extracting the counts from that source. We created one combined ACIS-S and ACIS-I spectrum for each source, fit thermal or non-thermal models to the combined spectra to create spectral models, and then computed PSF models using \textsc{marx}, simulating each source for a total of 5 Ms per frame and placing the source at its reference position. The sources with known proper motions were simulated at a shifted position in the 2018 frames, by adding the Gaia proper motions (multiplied by the 14.4 year baseline) to the 2004 reference positions. We then generated images of the \textsc{marx} PSF models and data, centered on the source reference position and using a bin size of 1/8th of an ACIS pixel. We limited ourselves to counts between 0.5 and 3.5 keV to minimize effects of the particle background and the wide PSF at higher energies. Because the PSF models were all placed at the reference position, the systematic offset between the data and the source model at the center of the image should now give us the offset of the frame with respect to the reference frame. 

The FoM is created by incrementally shifting the PSF model along the RA and DEC directions in steps of 1/8th ACIS pixel, at each step calculating the Poisson probability of the data given the model. The Poisson probability is only calculated for pixels within a circular aperture. The radius of this aperture should be large enough that sufficient counts be available for accurate centroid determination, but not so large that the FoM is significantly contaminated by background features. By calculating the frame offsets with varying aperture size, we determined that the smallest `stable' radius is around $3.5$ ACIS pixel radius. Once the FoM is constructed, a 2-D Gaussian function is fit to it to find the centroid and thus determine the offset of the source with respect to its reference position. To determine whether the reference positions estimated by \textit{wavdetect} are accurate, we first constructed the FoM for the sources in the reference frame. For some of the sources, we find an offset between the \textit{wavdetect} reference position and the FoM reference position. In these cases, we repeated the \textsc{marx} simulations for that source using the new FoM reference position. 

Once we verified that the FoM-determined position of each individual source in the reference frame is equal to its reference position, we added all the data and PSF models for each ObsID, and constructed a coherent FoM out of the summed data and summed model. From this coherent FoM we then obtain the offset of the frame with respect to the reference frame.

The expected $1\sigma$ statistical uncertainty in the RA and DEC directions is $\sim \theta_{\rm PSF} /N^{1/2}$, which varies from $13$ to $41$\,mas per frame. The excess systematic error seems to depend primarily on the energy bandwidth and the aperture size used to calculate the Poisson probability at each pixel of the FoM map. In order to estimate this systematic error, we calculated the offsets of each frame with varying aperture radii, varying from $2.5$ to $6$ ACIS pixel radius. Based on this, we estimate the $1\sigma$ systematic uncertainty to be $20$\,mas.

\subsection{PSR position}
\label{sec:pos}

\begin{figure}
\centering
\hspace*{-5mm}\includegraphics[width=3.7 in]{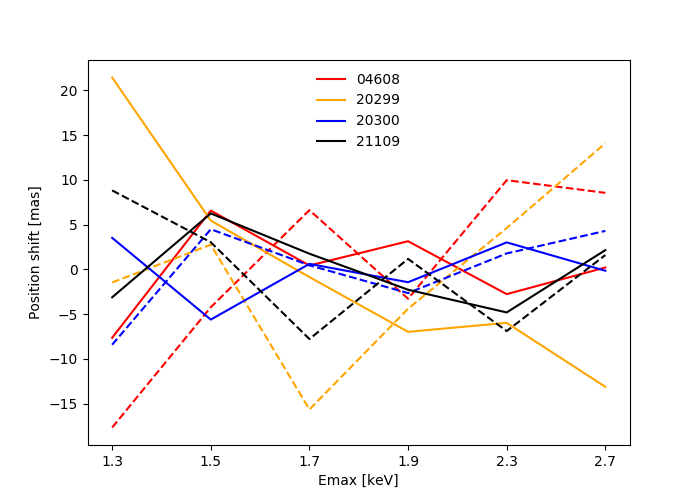} \\
\caption{Positional shift in RA and DEC of the PSR position in different frames as a function of the upper energy bound. The lower bound of the energy range was kept constant at $0.5$\,keV. Solid lines indicate RA shifts, dashed lines indicate DEC shifts. }
\label{fig:stat}

\end{figure}

To determine the position of the PSR in each frame, we proceeded as for the frame offsets. We extracted the PSR spectrum using a circular region with a $1.5^{\prime\prime}$ radius, then applied an aperture correction using the CIAO tool \textit{arfcorr}. We fit a BB+PL model to the PSR spectrum, and used the resulting model fit to simulate the PSR with \textsc{marx}. 

The PSR is lightly affected by pileup in all observations. Using the CIAO tool \textit{pileup\_map}, we estimate the pileup fraction to be $\approx 8\%$ in the brightest pixel of the reference observation, and $\approx 7\%$ in the brightest pixel of the 2018 observations, dropping down to a few percent when averaged over the $1.5^{\prime\prime}$ radius extraction region. No reference source is measurably affected. We have estimated how pileup affects our results by folding the pulsar PSF simulations through the \textit{marxpileup} tool. We determined the PSR positions in each frame using both the normal PSF and the piled up PSF. The pulsar positions change by approximately $\sim 0.3 \sigma$ between the two PSFs - a shift of low significance, indicating pileup is unlikely to affect our results. We also could not detect any systematic bias in the PSR position, comparing results with the piled and unpiled PSFs.

The error on the PSR position is more likely dominated by the surrounding bright, time-variable PWN. We have attempted to minimize the effect of PWN photons by using a narrower low energy band, thereby including most of the soft, thermal PSR photons but excluding some of the harder power law emission of the PWN. Of course, using too narrow an energy band decreases the number of available photons and therefore increases the statistical uncertainty of the PSR position. Figure \ref{fig:stat} shows how the PSR position shifts in RA and DEC for the four frames, as a function of the upper cutoff to the energy band (the lower bound was held fixed at $0.5$\, keV). The position shifts by $\sim 15$\,mas. We chose an energy band of $0.5-1.7$\,keV as a balance between systematic and statistical errors, but it is inevitable that our choice of energy band will have some small effect on the PSR position.

A similar consideration applies for the aperture size that is used to construct the FoM. For the pulsar, we chose a comparatively small aperture size of 2 ACIS pixel radius, in order to minimize the number of PWN photons while retaining sufficient photons within the aperture for accurate statistical determination of the PSR centroid. Calculating the PSR position with different aperture sizes shows that the position estimate shifts by a similar amount (1$\sigma$ uncertainty of $\sim 20$\, mas) as a function of aperture radius. Combining these energy cut and aperture radius effects, we estimate the total $1\sigma$ systematic uncertainty to be $\sim30$\,mas. The statistical uncertainties were computed from the pulsar aperture counts (as for the frame offsets) and range from $8$ to $14$\,mas. 

\subsection{Proper Motion}

The results of the proper motion analysis and associated $1\sigma$ uncertainties per frame are listed in Table \ref{tab:pm:shifts}. We calculated the uncertainty of the proper motion in each frame by adding 6 terms in quadrature: the statistical uncertainties of the location of the reference star grid in the individual frame and the reference frame, the systematic uncertainty of the frame shift, the statistical uncertainties of the PSR location in the individual frame and the reference frame, and the systematic uncertainty of the PSR frame location. Averaged over all 3 frames, we find a proper motion $\mu_{\rm RA} = 12 \pm 2 $\,mas\,yr$^{-1}$ and $\mu_{\rm DEC} = -1 \pm 2$\,mas\,yr$^{-1}$, for a total motion $\mu=12 \pm3$\,mas\,yr$^{-1}$.

Given the small angular velocity of J1709, local Galactic motions add a non-negligible contribution to the true proper motion of J1709 with respect to its local environment. This proper motion is a function of 1) the differential rotation of objects orbiting in Galactic disk of the local Galactic neighborhood, as described by the Oort constants A, B, C and K, 2) the motion of the Sun with respect to its local standard of rest, described by the vectors U, V, and W, 3) the Galactic coordinates $b$ and $l$ of the object, and 4) the distance to the object. Using equations 1 and 2 of \citet{Bovy2017}, the U, V, and W values of \citet{Schoenrich2010}, and assuming $d=2.3$\,kpc, we find that the proper motion induced by local motions is $\mu_{\rm RA, local} =-0.3$\,mas\,yr$^{-1}$ and $\mu_{\rm DEC, local} = -1.7$\,mas\,yr$^{-1}$. These numbers will increase by $\approx 10\%$ at $d=2$\,kpc, or decrease by $\approx 20\%$ at $d=3$\,kpc, well below the 1$\sigma$ uncertainty on J1709's proper motion in either case.

Correcting for the local proper motion shifts the true proper motion vector to $\mu_{\rm RA} = 13 \pm 2 $\,mas\,yr$^{-1}$, and $\mu_{\rm DEC} = 1 \pm 2$\,mas\,yr$^{-1}$, for a total motion of $\mu = 13 \pm 3$\,mas\,yr$^{-1}$ at a PA of $86 \pm 9 \degree$. The final proper motion vector, extrapolated backwards for the $1.7 \times 10^{4}$\,yr spindown age, is shown in Figure \ref{fig:largescale}. At $d = 2.3$\,kpc, our measured proper motion corresponds to a 2-D velocity $v_{\rm t} = 136 \pm 30$\,km\,s$^{-1}$. This is larger than the scintillation velocity estimate $v_{\rm t} \approx 90$\,km\,s$^{-1}$ \citep{Johnston1998}, suggesting that the scattering screen dominating the scintillation is closer than halfway to Earth or that the pulsar distance is $<2$\,kpc.

\begin{table}
\centering
\caption{Frame shifts and proper motions shown with 1$\sigma$ combined statistical and systematic confidence intervals. The reference frame, ObsID 4608, is marked with an asterisk.}
\begin{tabular}{cccccccc}
\hline \hline
ObsID & Shift RA & Shift DEC & $\mu_{\rm RA}$ & $\mu_{\rm DEC}$ \\  
& [$10^{2}$ mas] & [$10^{2}$ mas] & [mas yr$^{-1}$] & [mas yr$^{-1}$] \\ \hline
4608* & $ 0.1 \pm 0.1 $& $ 0.1 \pm 0.2 $ &\\ 
20299 & $ 2.0 \pm 0.3 $ & $0.2 \pm 0.2 $& $12 \pm3 $ & $-2 \pm 3$ \\
20300 & $ 1.6 \pm 0.3 $ & $0.4 \pm 0.3 $& $11 \pm 4$ & $-1 \pm 3$ \\
21109 & $ 2.1 \pm 0.4 $& $ 0.6 \pm 0.3 $ & $13 \pm 4$ & $1\pm 3$ \\ \hline

\end{tabular}

\label{tab:pm:shifts}

\end{table}

\section{Spectral analysis}

\subsection{Photon Index Map}

\begin{figure}

\centering
\includegraphics[width=0.5\textwidth]{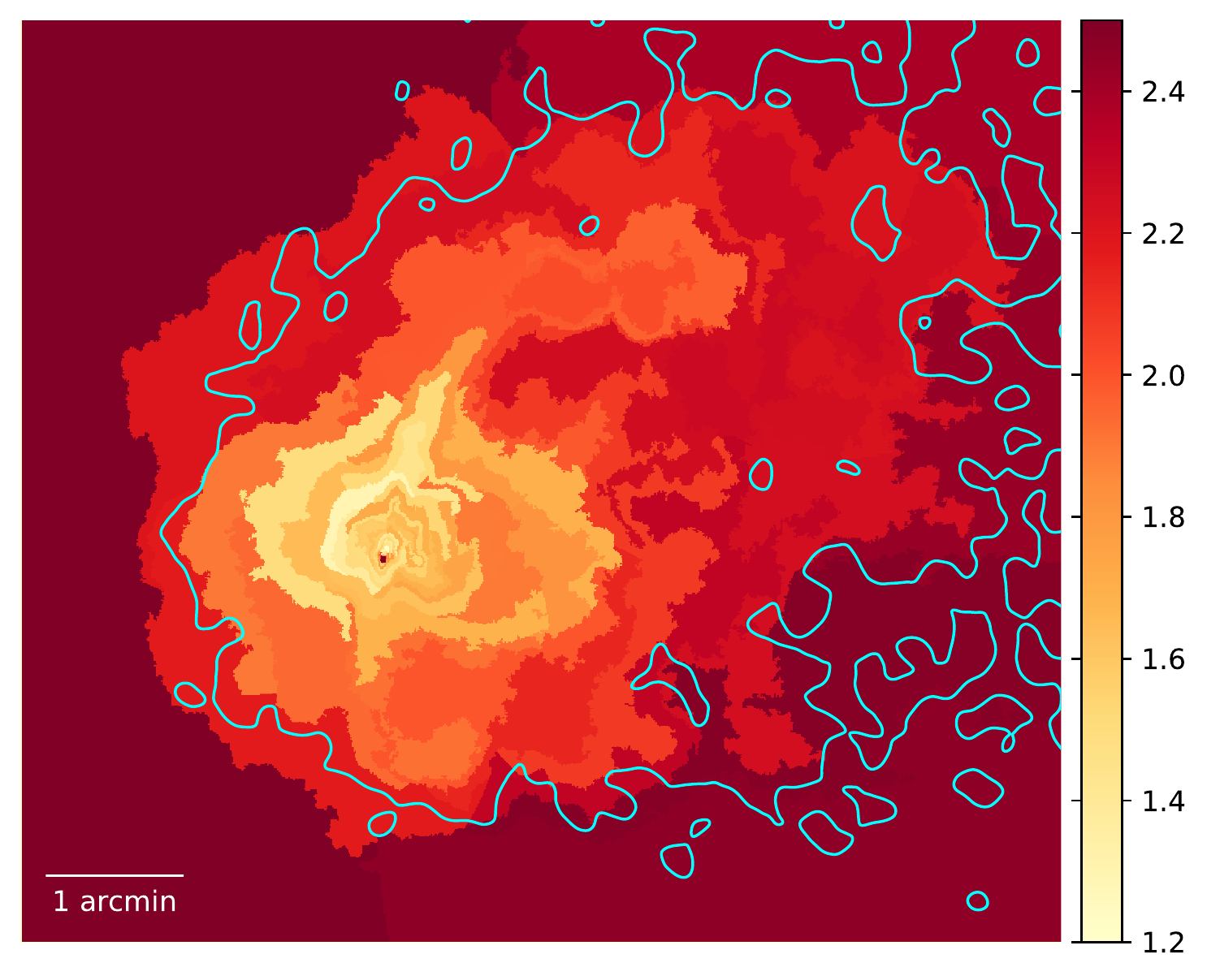} 
\caption{Photon index map of J1709. Spectral bins were defined following surface brightness contours. To the spectrum extracted from each bin, we fit a PL model with $N_H$ set to $5 \times 10^{21}$\,cm$^{-2}$. The typical 90\% error is $0.11$. The edge of the wind bubble (see Figure \ref{fig:bowshock}) is shown as the cyan contour. }
\label{fig:gammamap}
\end{figure}

We have constructed a photon index map of J1709 and its immediate surroundings by making bins with the \textit{contbin} algorithm \citep{Sanders2006}, often used to map the thermal properties of the intracluster medium in galaxy clusters. This algorithm constructs bins of approximately equal S/N, with bins following the surface brightness contours of the input data. We analysed a region of $450\times400^{\prime\prime}$, covering the PSR, PWN, the swept-back jets, and some of the diffuse X-ray emission trailing the system. Using a S/N of 25, we obtained 96 bins for which we extracted spectra using the CIAO tool  \textit{specextract}. We used the ACIS blank-sky files as the background. For each region, we created a combined ACIS-S and a combined ACIS-I spectrum using \textit{combine\_spectra}. We then fit an power law model to these spectra, setting the Galactic column density of $N_H$ to the value determined by \citet{Romani2005}, $N_H = 5 \times 10^{21}$\,cm$^{-2}$. The resulting map is shown in Figure \ref{fig:gammamap}.

In general, the photon index in the innermost regions is comparable to what is observed in other bright, spatially resolved PWNe  \citep{Kargaltsev2017}, albeit on the harder side. The spectrum softens with increasing distance from the PSR because of synchrotron losses. On the eastern side of the PWN equatorial outflow, where the torus is compressed by ram pressure, we observe a few bins where where the spectrum appears to be harder than the rest of the PWN, $\Gamma \approx 1.3$, and an inner arc on the eastern side appears even harder. These bins wrap around towards the jet which, in the north, also reveals particularly hard emission. The connection with the jets may explain the non-monotonic increase in $\Gamma$ with distance form the PSR. We also see a clear division between the emission inside and outside the pulsar wind bubble, with the emission inside the bubble being harder than the $\Gamma \approx 2.5$ background.

\subsection{Spectral fits of specific features}

\begin{figure}
\centering
\includegraphics[width=3.4 in]{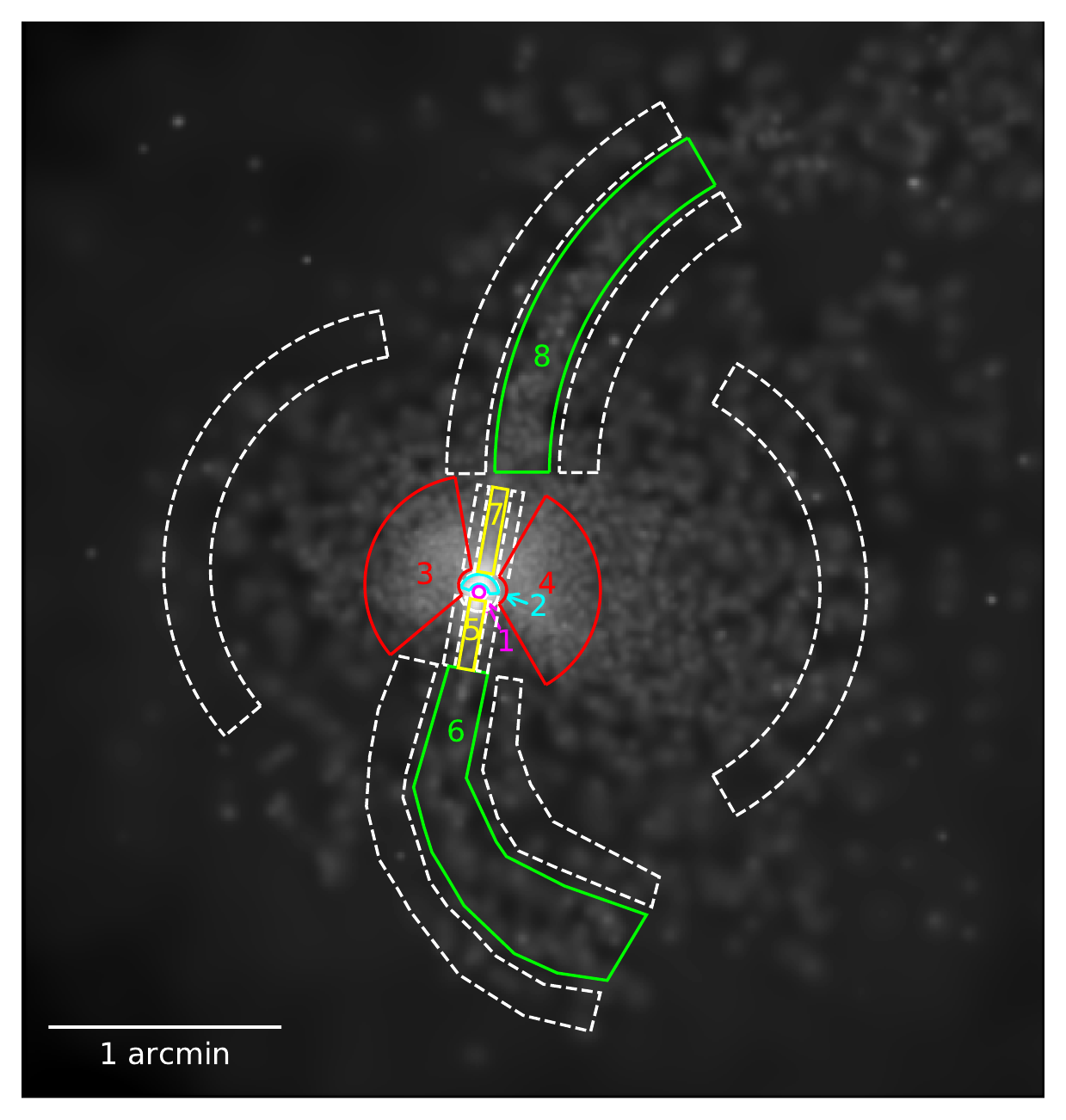} \\
\caption{Adaptively smoothed image of J1709, showing spectral extraction regions with solid lines and regions used for background subtraction with white dashed lines.}
\label{fig:specregs}

\end{figure}

\begin{table*}
\caption{Spectral fit results of regions shown in Figure \ref{fig:specregs}, with $0.5-7$ keV unabsorbed fluxes for each spectral model, and 90\% confidence intervals for all parameters. The Galactic absorption was held fixed at $N_H = 5 \times 10^{21}$\,cm$^{-2}$. The PSR region was fit with a composite blackbody + power law model, all other regions were fit with a power law model. }
\smallskip
\centering 
\begin{tabular}{l l l l l l } 
\hline\hline 
Nr & Comp		& Counts & $f_{\rm 0.5-7 keV}$ & kT/$\Gamma$ & $\chi^2$/DoF \\
&  &  & [erg\,cm$^{-2}$\,s$^{-1}$] & [keV]/ \\ \hline
1 & PSR (BB) & $15551 \pm 129$ & $(4.13 \pm 0.16) \times 10^{-13}$ & $0.180 \pm 0.004$ & $240/226$\\
1 & PSR (PL) &  " & $ (2.17 \pm 0.09 )  \times 10^{-13}$  & $1.46 \pm 0.09$ & " \\ 
2 & Torus & $5371 \pm 75$ & $ (1.26 \pm 0.03) \times 10^{-13}$ & $1.46 \pm 0.05$ & 216/211 \\
3 & PWN E & $5283 \pm 82 $&  $ (1.25 \pm 0.05) \times 10^{-13}$ & $1.49 \pm 0.06$ & 235/254 \\
4 & PWN W & $5627 \pm 87 $ & $ (1.35 \pm 0.05) \times 10^{-13}$ & $1.61 \pm 0.06$ & 184/188 \\ 
5 & Inner jet & $1657 \pm 49 $ & $ ( 3.9 \pm 0.2) \times 10^{-14}$ & $1.41 \pm 0.11$ & 159/185 \\
6 & Outer jet & $594 \pm 65$ & $ ( 1.4 \pm 0.4) \times 10^{-14}$  & $1.47\pm 0.53$ & 107/105\\
7 & Inner c/jet & $1242 \pm 54$ & $(2.9 \pm 0.3) \times 10^{-14}$&  $1.23 \pm 0.18$ & 129/195\\
8 & Outer c/jet & $1211 \pm 65$ & $(2.8 \pm 0.2) \times 10^{-14}$ & $1.33 \pm 0.24$ & 119/126\\

\hline 

\hline
\end{tabular}
\label{table:fits} 
\end{table*}

In addition to the photon index map, we have extracted the spectra from several individual features, which are shown in Figure \ref{fig:specregs}. For these fits we subtracted a local background spectrum, rather than using an ACIS `blank-sky' background. For the PSR, we assume that the emission is generated by a combination of thermal emission from the surface, as well as power law emission from the magnetosphere. We therefore fit a composite blackbody + power law model to the PSR spectrum. For the other regions, we assume a simple synchrotron power law spectrum. The Galactic absorption is again set to $N_H = 5 \times 10^{21}$\,cm$^{-2}$. The results of these fits are listed in Table \ref{table:fits}. These fits show similar trends to those observed in Fig. \ref{fig:gammamap}, including a clear asymmetry between the eastern and western side of the PWN. Additionally, we note that for the inner jet region, the ACIS-S and ACIS-I combined spectra show different values of $\Gamma$ when fit individually, $\Gamma=1.55 \pm 0.22$ and $\Gamma=1.33\pm 0.14$ respectively. The difference is barely significant, but it may be the result of month-timescale jet variability, as seen in Figure \ref{fig:spots}. 

We can use these spectral measurements to estimate the magnetic field strength in the synchrotron-emitting PWN components. For an optically thin region filled with relativistic electrons and magnetic field emitting synchrotron radiation
\begin{equation}
\label{eq:syncB}
B = 46 \left[ \frac{J_{\rm -20}(E_1,E_2) \sigma } {\phi}  \frac{C_{1.5-\Gamma}(E_m, E_M)}{C_{2-\Gamma}(E_1,E_2)}\right] ^{2/7} \mu G
\end{equation}
where 
\begin{equation}
C_q(x_1,x_2) = \frac{x_2^{q} - x_1^q}{q}.
\end{equation}
$J_{\rm -20}(E_1,E_2) = 4 \pi f_{\rm -20}(E_1, E_2) d^{2}/ V$ is the observed emissivity (in $10^{-20}$\,erg\,s$^{-1}$\,cm$^{-3}$, between $E_1$keV and $E_2$keV), $\sigma=w_B/w_e$ is the magnetization parameter, $\phi$ the filling factor, and $E_m$ and $E_M$ the minimum and maximum energies of the power law $\Gamma$ synchrotron spectrum, in keV. 

For the inner jet and inner counterjet, we use  the fluxes from Table \ref{table:fits}, and estimate the jet volume as a cylinder inclined $\sim37\degree$ from the sky plane. If we assume $\sigma=1$, a distance $d=2.3$\,kpc, $\phi=1$ and that the photon spectrum extends from $1\,{\rm eV} - 1 \,{\rm MeV}$, we obtain $B=43\,\mu$G for the inner jet and $B=40\,\mu$G for the inner counter-jet. 

For the PWN, we divided regions 3 and 4 into inner and outer parts with equal radius. We estimated the volume of these regions as spherical shells projected along the line of sight. Using the same assumptions as for the jet and counterjet, we estimate that $B=27\,\mu$G for the inner PWN, and $B=15\,\mu$G for the outer PWN. These are rough estimates only, as the the jet clumping suggests $\phi < 1$ and multi-wavelength data at present provide little constraint on the full spectral range.

\section{The variable jets and PWN}

\begin{figure}

\hspace*{-5mm}
\includegraphics[width=.53\textwidth]{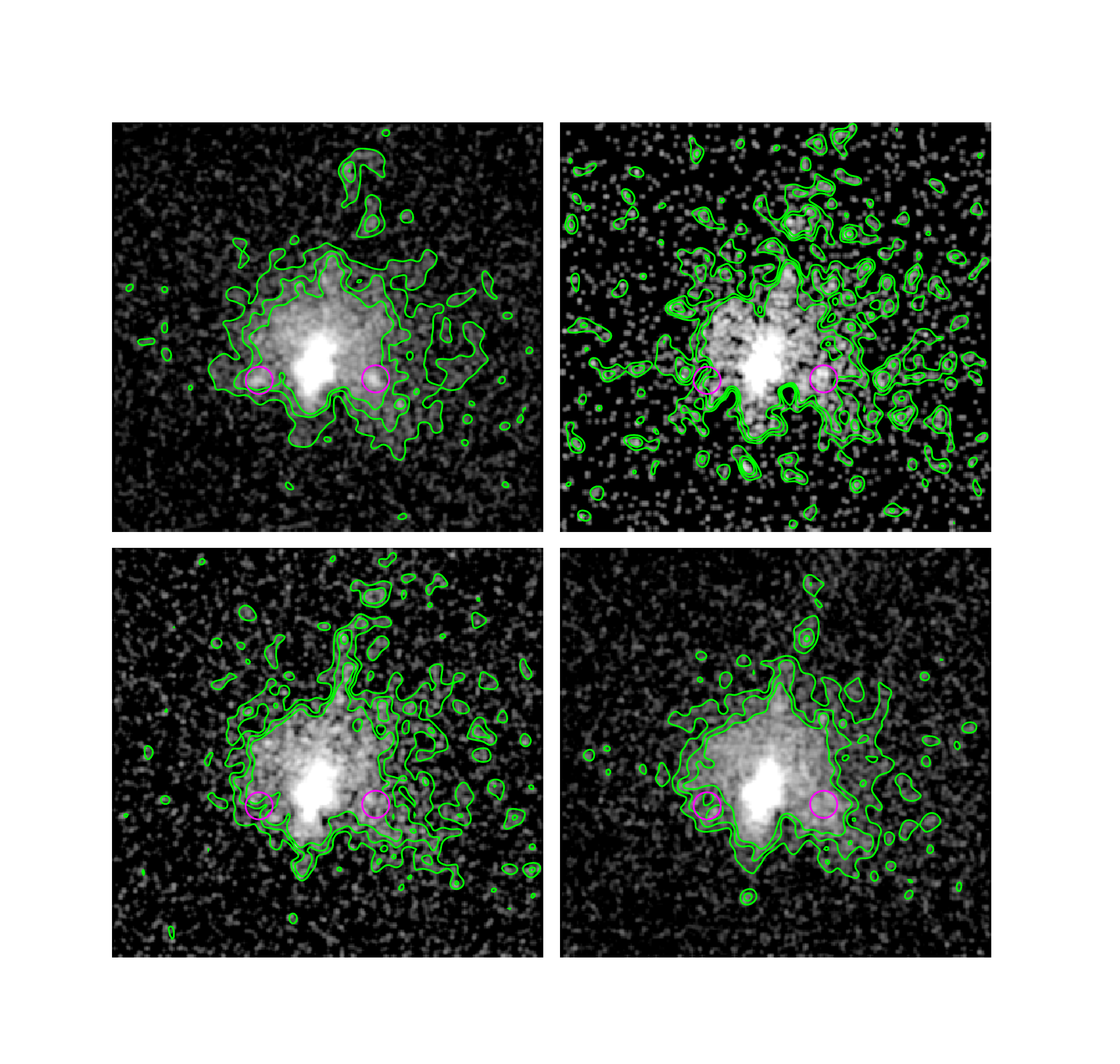} 
\vspace*{-5mm}
\caption{Exposure-corrected images of J1709: epoch 3 (top left), 4 (top right), 5 (bottom left), and 6 (bottom right). Surface brightness contours are shown in green, to emphasize the jet variability between epochs. The PWN spots (primarily visible in epoch 3) are marked with magenta circles.}
\label{fig:spots}

\end{figure}

We have divided the total \textit{CXO} data set of J1709 in 6 different epochs: 1) 2000 [15\,ks], 2) 2004 [99\,ks], 3) Jun 2018 [230\,ks], 4) Oct 2018 [53\,ks], 5) Jan-Feb 2019 [173\,ks], 6) May-Jun 2019 [216\,ks]. Closer inspection of the system at these different epochs reveal some interesting structural variation. In epoch 3, two bright spots appear in the outflow, roughly equidistant east and west from the PSR. Exposure-corrected images for epochs 3, 4, 5 and 6 are shown in Figure \ref{fig:spots}. The spots are statistically significant, with a S/N ratio of 5.6 for the eastern spot and 6.0 for the western spot. Unfortunately, the subsequent epoch 4 is limited by poor count statistics as the observations were interrupted by \textit{CXO} going into safe mode. The western spot appears still visible at the same location in this epoch, at a S/N ratio of 3.6. 

We have extracted spectra from the spots, but a detailed analysis is difficult with the limited number of counts and low contrast with the surrounding PWN. For the eastern spot, a background-subtracted spectrum (using a background annulus directly around the spot) results in a spectrum with $68$ counts between 0.7 and 6\,keV. Fitting an absorbed PL with $\rm{N_H} = 5 \times 10^{21}$\,cm$^{-2}$ yields a photon index of $1.35 \pm 0.67$. For the western spot, we obtain $97$ counts and a photon index of $1.15 \pm 0.46$. While nominally harder than the surrounding torus/outflow emission, the difference is not statistically significant.

\begin{figure}

\centering
\includegraphics[width=0.5\textwidth]{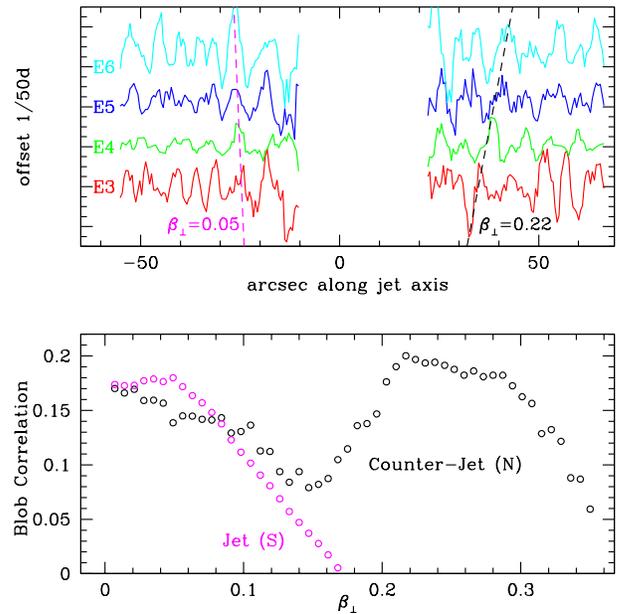} \\
\caption{\textit{Top panel}: Counts profiles along the jet axis for epochs 3, 4, 5, and 6. These profiles are offset vertically by one tick mark/50days between epochs. The x-axis direction is south (jet) to north (counterjet). \textit{Bottom panel}: Result of the cross-correlation for jet and counterjet, suggesting blob motions $\beta_\perp \approx 0.05$ for the jet, and $\beta_\perp \approx 0.22$ for the counterjet.}
\label{fig:jetcc}

\end{figure}

The jets of J1709 also seem to show variability in different epochs, somewhat reminiscent of the variable jet in the Vela Pulsar \citep{Pavlov2003}. In particular, there seems to be some side-to-side instability in the jet axis in epoch 5. However, on larger scale the jets are remarkably narrow and curve smoothly, apparently swept back by flow within the pulsar wind bubble. The limited jet spread implies some source of continuous re-colimation (possibly magnetic hoop stress). We also attempted to analyze intra-epoch jet blob motion by using a cross-correlation routine to compare counts profiles along the jet axis at different epochs, as shown in Figure \ref{fig:jetcc}. This showed little lagged correlation in the southern (jet) source, consistent with $\beta_\perp = v_\perp/c < 0.05$. For the northern (counter-)jet, a broad correlation peak suggests $\beta_\perp \approx 0.22$. However no epoch showed bright stable blobs, and we fear that the significance of these  correlations is difficult to quantify. Thus we will not use these $\beta$ values to draw physical conclusions.

\section{Discussion and Conclusions}

Armed with a direct proper motion measurement, we can re-evaluate the association of PSR~J1709$-$4429 with SNR~G343.1$-$2.3. The $\sim 25^{\prime}$ radius radio arc has been proposed by \citet{Bock2002} to be the interaction of the blastwave with a molecular cloud, with the balance of the shell $3-4\times$ larger, faint or invisible in a less-confined region to the east. These authors require a proper motion $35$\,mas\,yr$^{-1}$ to the SE to bring the PSR to its present location in the spindown age. In contrast, our proper motion is $\sim 3\times$ smaller, nearly due east, so birth at the center of the radio arc is excluded. On the other hand, few pulsars as young as 17\,kyr lack parent supernova remnants, and with the low proper motion guaranteed for J1709, G343.1$-$2.3 is the only candidate in the vicinity.

If, instead, the birthsite is 17\,kyr back along our proper motion vector (base of the magenta arrow in Fig.\,\ref{fig:largescale}) the explosion was $\sim 8^\prime$ south of the center of the radio arc. One might infer that the pulsar progenitor traveled south toward this point, evacuating a cavity {\it via} a strong stellar wind. Explosion into this cavity led to the offset (but surprisingly symmetric) radio arc seen today. In response to this explosion the pulsar now travels east at $\sim 90 \degree$ to its progenitor's motion. 

We have noted that the combined radio/X-ray/TeV trail matches well with the proper motion axis and terminates in a rounded front centered on the pulsar. This can be interpreted as a trailed PWN. In this picture it is natural to associate the TeV peak \citep{Hoppe2009} with `remnant' electrons marking the pulsar's rapid initial spindown, while the X-ray trail following the torus and jets represent younger, higher energy electrons. However, we would require $\sim 5\times\tau_c \approx 85$\,kyr (cyan arrow in Figure \ref{fig:largescale}) for travel from the TeV peak. An alternative origin for the TeV emission is G343.1$-$2.3-generated cosmic rays interacting with the dense gas marking the remnant's western boundary at the radio arc. Indeed the lobe of TeV emission extending north into the radio arc supports this interpretation for at least part of the flux. However the extension toward the pulsar argues that at least some of the TeV emission is associated with J1709. The faint lobe to the west may be insignificant or background from the Galactic plane. The improved spectral imaging expected with the Cherenkov Telescope Array may be able to address the origin of this TeV emission.

It is relatively common for young pulsars to have spindown ages $\tau_c$ in excess of their kinematic ages, e.g.\ indicating a large spin period at birth. There are only a few PSR for which the opposite has been suggested. Perhaps the most discussed example is the PSR~B1757$-$24/G5.4$-$1.2 association (`the Duck'), where the upper limit on the proper motion requires a true age of $>4\times \tau_c$ for the association to be real \citep{Zeiger2008}. One way to reconcile such age differences is late time growth of the magnetic field \citep{Blandford1988} or magnetic geometry changes that increase the braking torque within the past $\sim \tau_c$. A similar argument might apply to J1709, although one would require a large field near birth if the high energy deposition at the TeV maximum is to be explained. One could then speculate that the spindown torque oscillates, e.g.\,due to free precession on a timescale of a few $\tau_c$. If the pulsar really were born at the current location of the TeV peak at the edge of the radio shell, the required extreme asymmetry of the remnant expansion would be very difficult to understand. Of course, if the true age is $\geq 85$\,kyr, it is also easier for the parent SNR to have faded below visibility, and so the association with G343.1$-$2.3 might not be required in this scenario.

Interpreting the X-ray/radio/TeV horizontal band as a PWN trail, it is of interest to understand why this region extends $\theta_0 \sim 90^{\prime\prime}$ in front of the pulsar. One possibility is that this is a classical bow shock, set by momentum balance between the pulsar wind and the ram pressure of the external medium, with
\begin{equation}
\begin{aligned}
\theta_0 = & [{\dot E}/(4\pi c \rho v^2)]^{1/2}/d \\
= & 150^{\prime\prime}[{\dot E_{36}} {\rm \sin}^2i/n_p \mu_{\rm mas/y}^2]^{1/2} d_{\rm kpc}^{-2}.
\end{aligned}
\end{equation}
This applies when the pulsar motion is supersonic in the external medium. For such a large standoff J1709's low proper motion helps, but we would also need a low external density $n_p \approx 2.5 \times 10^{-3} (d/2.3{\rm kpc})^{-4} {\rm cm^{-3}}$. Although such low densities can occur in stellar wind bubbles or supernova remnants, but with long cooling times the temperatures and pressures in such cavities would be too high for the pulsar motion to be supersonic.

The standoff may instead be controlled by energy balance for a sub-sonic pulsar. Here the pulsar wind bubble expands in a Sedov-like solution to
\begin{equation}
\begin{aligned}
R(t)= & [1.3 {\dot E}t^3/\rho ]^{1/5}\
= & 0.6({\dot E_{36}}/n_p)^{1/5} t_3^{3/5}{\rm pc},
\end{aligned}
\end{equation}
where $t_3$ is a characteristic energy deposition time, in units of $10^3$\,yr. The leading edge of this wind bubble holds steady at the present standoff when the expansion velocity
\begin{equation}
v(t)= {\dot R}(t) = 352 ({\dot E_{36}}/n_p)^{1/5} t_3^{-2/5}{\rm km\,s^{-1}}
\end{equation}
balances the pulsar motion, with ratio $R/v = 1670 t_3 {\rm yr}$. Observationally, $R/v\approx 7,200$\,yr so we infer $t_3\approx 4.3$, that is to say that the effective confinement time of the pulsar wind is $\sim 4,300$\,yr with the relativistic $e^\pm/B$ plasma venting behind the pulsar, to the west. For this timescale and the pulsar spindown power we infer a density $n \approx 0.084 d_{2.3}^{-5} {\dot E}_{36} t_3^3\,{\rm cm^{-3}} \approx 27\,{\rm cm^{-3}}$ from the wind bubble standoff with the nominal confinement time, or $\approx 0.34\,{\rm cm^{-3}}$ for more efficient wind venting with $t_3\sim 1$. Given the proximity to the Galactic plane and evidence for molecular gas west of G343.1$-$2.3, a large density does not seem unreasonable if the pulsar is {\it not} embedded in the SNR, although it would likely require some pulsar motion out of the plane of the sky to escape to a dense external medium. This wind venting is appealing as it provides a natural source of the backflow that sweeps the jets back toward the pulsar tail, with speeds several times $v_{\rm PSR}$. In addition, the venting PWN trail will sweep partly cooled $e^\pm$ to the west, perhaps even beyond the pulsar birthsite to the TeV peak, mitigating the disagreement with the characteristic age.

The torus and jets themselves can be related to the proper motion. \citet{Romani2005} fit the torus/jet axis at a position angle $164\degree$, some $77\degree$ from the proper motion axis. In addition the relativistic torus fit implies the (southern) jet axis projects $\sim 37\degree$ out of the plane of the sky. If we assume a proper motion within the plane of the sky, then the jet (pulsar spin) axis lies $\sim 80^\circ$ from the proper motion vector. In \citet{Ng2007} it was found that when both spins and proper motions are introduced by a natal kick, low-velocity objects have the highest probability of large spin-velocity angles.

Note that the southern jet is pointing against the venting wind flow and is truncated. If the 3D space velocity has a component out of the plane of the sky, then this jet would be pointing somewhat closer to the forward direction. This should enhance the back-flow asymmetry with the northern counter-jet which sweeps back and fades smoothly into the PWN trail. Thus 3-D relativistic MHD simulations following the polar field lines of the jet and counter-jet might be able to constrain the 3-D space motion, and by constraining the venting flow speed can determine the upstream density and the distance traveled by the TeV emitting $e^\pm$.

\bigskip
We conclude the PSR J1709 was not born at the center of SNR~G343.1$-$2.3, but instead the explosion occurred somewhere between its present position and the TeV peak. If the true age is $\sim \tau_c$, then the SNR association is plausible, but requires progenitor motion, substantial expansion asymmetry, a birth kick nearly orthogonal to this asymmetry and substantial flow of the PWN $e^\pm$ (if these cause the TeV emission).
Alternatively, if born at the TeV peak $\sim 5 \tau_c$ ago, then the required explosion symmetry is even more extreme. This places the association in doubt, but does allow more time for a hypothetical older birth SNR to fade.

Tunneling into a relatively dense medium, the pulsar is evacuating a wind-blown trail, whose relativistic shocked $e^\pm$/B fluid is venting to the west. The bubble scale suggests a bulk outflow speed $> 4\times v_{\rm PSR}$. This flow is graphically displayed by the jets which mark the polar axis field lines of the PWN outflow. These remain relatively well collimated, but are swept to the west on a timescale shorter than the pulsar crossing time. The high spindown power (and relatively low pulsar velocity) guarantee that the jets and torus are not crushed as in some older pulsar bow shocks. In this picture the jet sweep back is not a direct effect of the pulsar motion, but rather a secondary effect of its (motion-induced) tunneling into a dense medium with backflow escaping behind the pulsar and sweeping the jets along with this venting wind. Simulations are needed to see if this picture is viable and to further constrain the flow speed.

\acknowledgements
MdV and RWR were supported in part by NASA grant G08-19050A, through the Smithsonian Astrophysical Observatory. GGP and BP were supported by NASA grant G08-19050B. OK was supported by NASA  grant GO8-19050C and ADAP grant 80NSSC19K0576. Support for this work was provided by the National Aeronautics and Space Administration through Chandra Award Number GO8-19050 issued by the Chandra X-ray Observatory Center, which is operated by the Smithsonian Astrophysical Observatory for and on behalf of the National Aeronautics Space Administration under contract NAS8-03060.

\vspace{5mm}

\bibliography{J1709}{}
\bibliographystyle{aasjournal}



\end{document}